\documentclass[11pt]{article}
\usepackage[margin=1.05in]{geometry}
\usepackage[T1]{fontenc}
\usepackage{lmodern}
\usepackage{amsmath,amssymb,amsthm,mathtools}
\usepackage{microtype}
\usepackage[colorlinks=true,citecolor=blue!55!black,linkcolor=blue!45!black,urlcolor=blue!55!black]{hyperref}
\usepackage{xcolor}
\usepackage{enumitem}
\setlist{nosep,leftmargin=*}
\allowdisplaybreaks[2]
\newtheorem{theorem}{Theorem}[section]
\newtheorem{lemma}[theorem]{Lemma}

\newtheorem{corollary}[theorem]{Corollary}
\newcommand{\E}{\mathbb E}
\newcommand{\Pbb}{\mathbb P}
\newcommand{\Var}{\operatorname{Var}}
\newcommand{\Inf}{\operatorname{Inf}}

\newcommand{\Norm}[1]{\left\lVert#1\right\rVert}
\newcommand{\R}{\mathbb R}
\newcommand{\cX}{\mathcal X}
\newcommand{\cY}{\mathcal Y}
\newcommand{\ind}{\mathbf 1}
\title{Polynomial Sample Complexity of Noninteractive Simulation over Finite Alphabets}
\author{Haoran Wang\\\texttt{whr.hrwang@gmail.com}}
\date{}
\begin{document}
\maketitle
\begin{abstract}
We study noninteractive simulation from a finite correlated source. Alice and Bob observe independent copies of the source and generate outputs without communication. We give a quantitative construction for arbitrary finite output alphabets: for every fixed source, a joint law that is simulable with arbitrarily many samples can be approximated using a number of samples polynomial jointly in the output alphabet sizes and the inverse accuracy. We first study binary outputs. A one-sided Wright--Fisher comparison gives a source-dependent explicit exponent and a finite-head Gaussian-threshold representation. We then handle arbitrary finite output alphabets through averaged influence regularity, simplex-preserving Gaussian moment lifting, and a common Gaussian projection for all correlated-prefix states. The sample exponent is independent of the output alphabet sizes. For a concrete biased binary source we certify an explicit exponent below $19$.
\end{abstract}

\section{Introduction}
Noninteractive simulation asks whether two parties observing independent copies of a correlated source can generate a prescribed joint distribution without communication. Private randomness is allowed. The sample complexity question is quantitative: if a target belongs to the closure of the simulable laws, how many source samples suffice to approximate it to a given accuracy?

Maximal correlation and hypercontractivity provide basic constraints on noninteractive simulation \cite{KA16}. Ghazi, Kamath, and Sudan established decidability for binary targets \cite{GKS16}. De, Mossel, and Neeman extended decidability to arbitrary finite target alphabets \cite{DMN18}. Their work on Gaussian noise stability also gives computable finite-dimensional approximations \cite{DMN19}. Ghazi, Kamath, and Raghavendra improved the quantitative dimension bounds through Gaussian projection \cite{GKR18}. Heilman and Tarter subsequently proved an exact finite-dimensional reduction for correlated Gaussian sources \cite{HT24}. Shirani and Heidari studied Fourier descriptions of finite-output simulation regions for binary input sources \cite{SH23}. Salehi and Shirani studied noninteractive simulation from binary sources using Fourier-analytic methods, including lexicographic constructions for sources with uniform input marginals \cite{SS25}. The general correlation inequalities and invariance principles of Mossel, O'Donnell, and Oleszkiewicz \cite{MOO10}, and of Mossel \cite{Mos10}, form the analytic background.

We prove that the required sample count is polynomial jointly in the two output alphabet sizes and inverse accuracy, with an exponent depending only on the fixed source. Our proof for general outputs has three parts. First, a single set of input coordinates is revealed so that both conditioned output functions have small average low-degree influences. Each player then removes the remaining exceptional coordinates using only that player's own prefix. Next, a quantitative moment-lifting argument transfers the resulting simplex-valued functions to Gaussian space. The transfer matches all Hermite coefficients through a logarithmic degree while preserving the simplex constraint. Finally, a normalized Gaussian projection simultaneously compresses every conditional strategy. The projection is chosen once for all prefix states, so its dimension does not depend on their number.

We begin with binary outputs. A one-sided Wright--Fisher diffusion leaves Alice's influences unrestricted while Gaussian smoothing controls Bob's quadratic variation. This method builds on the use of stochastic embeddings in noise stability, particularly the renormalized Brownian-motion approach of Eldan, Mikulincer, and Raghavendra \cite{EMR23}. It yields an explicit sample exponent and a finite-head Gaussian-threshold representation. We then treat general output alphabets by a separate moment-lifting method. The two approaches address the same sample-complexity problem at different levels of generality. The binary argument is not needed for the general output theorem, but it proves a one-sided comparison with the explicit Gaussian function $J_\rho$. It also gives a one-dimensional Gaussian-threshold normal form and the exponent $A(q)=2q(4q-3)/(q-2)^2$. For the biased source in Section~4, the exact $L^2\to L^5$ contraction yields $A(5)=170/9$; the weaker $L^2\to L^4$ certificate already gives $A(4)=26$. By contrast, the general-output proof gives a larger source-dependent exponent. Its finite-source Gaussian-tail implementation costs at most $(km/\varepsilon)^{29}$ samples, while the head regularity step may require more. The advantage is that one exponent works jointly for all finite output alphabet sizes. The two methods therefore give different information about the same problem.

\section{Main results and notation}
Let $P$ be a probability distribution on finite alphabets $\cX\times\cY$, with marginals $\nu,m$. Letters of zero marginal probability are omitted. For a function $h:\cY\to\R$, write
\begin{equation}\label{eq:channel}
(Th)(x)=\sum_yP(y\mid x)h(y),\qquad \rho=\sup_{\substack{\E_mh=0\\\Norm{h}_{L^2(m)}=1}}\Norm{Th}_{L^2(\nu)}.
\end{equation}
Thus $\rho$ is the maximal correlation of $P$. No assumption of full support is made on the joint law. For $a,b\in[0,1]$, set
\begin{equation}\label{eq:M}
M_n(a,b)=\max_{\substack{f:\cX^n\to[0,1],\ g:\cY^n\to[0,1]\\\E f=a,\ \E g=b}}\E[f(X^n)g(Y^n)],\qquad M_\infty(a,b)=\sup_{n\ge0}M_n(a,b).
\end{equation}
The range $[0,1]$ accounts for private randomization. Throughout, $C_P$ denotes a finite constant depending only on the fixed source and may change from line to line. The notation $O_P(t)$ means a quantity bounded in magnitude by $C_Pt$. For $0<\rho<1$, these constants may depend on the two marginal alphabet sizes, the smallest positive marginal masses $\min_x\nu(x)$ and $\min_y m(y)$, the spectral separation $1-\rho$, and the exponent $q>2$ in Lemma~\ref{lem:hyper}. They are not asserted to be uniform as $\rho\uparrow1$ or as a positive marginal mass tends to zero. The cases $\rho=0$ and $\rho=1$ are treated separately. Let $\Phi,\varphi$ denote the standard normal distribution and density, and put $I(a)=\varphi(\Phi^{-1}(a))$. For $0\le r<1$, let
\begin{equation}\label{eq:J}
J_r(a,b)=\Pbb[Z\le\Phi^{-1}(a),\ W\le\Phi^{-1}(b)],\qquad \operatorname{Corr}(Z,W)=r,
\end{equation}
where $Z,W$ are standard Gaussian. Continuous extensions at $a,b\in\{0,1\}$ are understood.

We use the Efron--Stein decomposition $h=\sum_{S\subseteq[n]}h_S$ relative to the relevant product marginal. Its low-degree influence is $\Inf_i^{\le d}(h)=\sum_{S\ni i,\ |S|\le d}\Norm{h_S}_2^2$.

For $k\ge2$, write $\Delta_k=\{u\in[0,1]^k:\sum_{a=1}^k u_a=1\}$. For integers $k,m\ge2$, let $\mathcal S_n(P;k,m)$ be the set of joint laws generated by independent private-randomized maps from $\mathcal X^n$ to $[k]$ and from $\mathcal Y^n$ to $[m]$. The sets are nested in $n$. We write $\mathcal S_\infty(P;k,m)$ for the closure of their union. Define $N_P^{k,m}(\varepsilon)$ to be the least $N$ such that every law in $\mathcal S_\infty(P;k,m)$ lies within total-variation distance $\varepsilon$ of some law in $\mathcal S_N(P;k,m)$.

The following continuity estimate will also be used when passing from finite protocols to the closed simulation region.
\begin{lemma}\label{lem:Lipschitz}
For every $n$, including $n=\infty$, and all $a,a',b,b'\in[0,1]$,
\[
|M_n(a,b)-M_n(a',b')|\le |a-a'|+|b-b'|.
\]
\end{lemma}
\begin{proof}
Suppose $a<a'$. Given an admissible $f$ of mean $a$, set $f'=f+(a'-a)(1-f)/(1-a)$. Then $0\le f'\le1$, $\E f'=a'$, and $\E|f'-f|=a'-a$. If $0<a'<a$, set $f'=(a'/a)f$; the endpoint cases follow by the same formulas or by continuity. Since $0\le g\le1$, changing Alice's function changes $\E[fg]$ by at most $|a-a'|$. Do the same for Bob, and compare optimizers in both directions. The estimate is uniform in $n$, so it passes to the supremum defining $M_\infty$.
\end{proof}

\subsection*{Binary comparison and quantitative approximation}
The first theorem is a one-sided comparison: Alice may use an arbitrary Boolean-valued randomized strategy, while only Bob has to satisfy a low-degree influence hypothesis. This asymmetry is what permits the refined binary estimate.

\begin{theorem}\label{thm:comparison}
Suppose $0<\rho<1$. There is $q=q(P)>2$ and $C_P<\infty$ such that, for every $n,d\ge2$, every $f:\cX^n\to[0,1]$, and every $g:\cY^n\to[0,1]$ satisfying $\max_i\Inf_i^{\le d}(g)\le\tau$, the following holds for $0<\theta<1/4$, $0<\lambda\le1$, and $T_0\ge1$:
\begin{equation}\label{eq:main-comparison}
\E[fg]\le J_\rho(\E f,\E g)+C_P\left(\theta+e^{-T_0/2}+\frac{\lambda(1+T_0)}{\theta}+\frac{d\tau^{q/2-1}}{\lambda^q}+\frac{(d+1)\rho^{2d}}{\lambda^2}\right).
\end{equation}
There is no influence assumption on $f$. The same assertion holds with the two parties interchanged.
\end{theorem}

Choose $q>2$ so that both $T:L^2(m)\to L^q(\nu)$ and its adjoint $T^*:L^2(\nu)\to L^q(m)$ are contractions; the existence of such a $q$ is proved below. Define
\begin{equation}\label{eq:exponent}
A(q)=\frac{2q(4q-3)}{(q-2)^2}.
\end{equation}
\begin{theorem}\label{thm:sample}
For every fixed finite source $P$ with $0<\rho<1$ and every $\delta>0$, there is $C_{P,\delta}<\infty$ such that, for all $n\ge1$,
\begin{equation}\label{eq:rate}
\sup_{a,b\in[0,1]}\bigl(M_\infty(a,b)-M_n(a,b)\bigr)\le C_{P,\delta}n^{-1/(A(q)+\delta)}.
\end{equation}
Consequently, every binary target law in the closure of the laws simulable from $P$ can be approximated to total-variation error $\varepsilon$ using at most $C_{P,\delta}\varepsilon^{-(A(q)+\delta)}$ samples. A polynomial bound holds for every finite $P$, including the cases $\rho=0$ and $\rho=1$.
\end{theorem}

There is also a structural form of the theorem. For $k\ge0$, define
\begin{equation}\label{eq:Vk}
V_k(a,b)=\max_{\substack{u:\cX^k\to[0,1],\ v:\cY^k\to[0,1]\\\E u=a,\ \E v=b}}\E\bigl[J_\rho(u(X^k),v(Y^k))\bigr].
\end{equation}
\begin{corollary}\label{cor:head}
Under the assumptions of Theorem~\ref{thm:sample}, $V_k(a,b)$ increases to $M_\infty(a,b)$ uniformly in $(a,b)$. For every $\delta>0$,
\begin{equation}\label{eq:head-rate}
0\le M_\infty(a,b)-V_k(a,b)\le C_{P,\delta}(k+1)^{-1/(A(q)+\delta)}.
\end{equation}
Thus every asymptotically simulable binary law has an approximation consisting of a polynomial-size finite head, a Gaussian threshold tail, and private output randomization. In the limiting output region, the possible values of $\Pbb[U=V=1]$ at fixed marginals $(a,b)$ form exactly the interval
\begin{equation}\label{eq:region}
\bigl[a-M_\infty(a,1-b),\ M_\infty(a,b)\bigr].
\end{equation}
\end{corollary}

\subsection*{General finite output alphabets}
The next theorem removes the output-size restriction. Its proof uses a simplex-preserving transfer principle rather than the binary Gaussian comparison. The constants below depend only on the fixed source, even when the two output alphabets vary with the desired accuracy.

\begin{theorem}\label{thm:general}
For every finite source $P$ there are constants $C_P,A_P<\infty$, depending only on $P$, such that
\begin{equation}\label{eq:general-main}
N_P^{k,m}(\varepsilon)\le C_P\left(\frac{km}{\varepsilon}\right)^{A_P}\qquad(k,m\ge2,\quad0<\varepsilon<1/2).
\end{equation}
For $0<\rho<1$, every law in $\mathcal S_\infty(P;k,m)$ is within total-variation distance $\varepsilon$ of a law of the following form: Alice and Bob use the same fixed set $H$ of at most $C_P(km/\varepsilon)^{A_P}$ input coordinates as a head. Conditioned on their respective prefixes, they apply simplex-valued functions to $D$ pairs of $\rho$-correlated Gaussian variables, where $D\le C_P(km/\varepsilon)^4$. The Gaussian tail can be implemented by at most $C_P(km/\varepsilon)^{29}$ further samples of the original source.
\end{theorem}

\noindent The binary and general results have distinct proofs. The next two sections develop the binary comparison and its refined consequences. Sections~\ref{sec:vector-regularity}--\ref{sec:general-proof} establish the general result.

\section{Binary Gaussian comparison via a one-sided diffusion}\label{sec:binary-comparison}
We start with two estimates that are also useful for the general-output argument. The first follows from the strict maximal-correlation gap on a finite source. The second turns bounded one-coordinate increments into a Gaussian first-level estimate.

\begin{lemma}\label{lem:hyper}
For a finite joint source with $0<\rho<1$, there is $q>2$ such that
\begin{equation}\label{eq:hyper}
\Norm{Th}_{L^q(\nu)}\le\Norm{h}_{L^2(m)},\qquad \Norm{T^*k}_{L^q(m)}\le\Norm{k}_{L^2(\nu)}.
\end{equation}
Both inequalities hold for every tensor power, with the same exponent and norm bound.
\end{lemma}
\begin{proof}
We prove the first estimate. The other follows by exchanging the marginals. Let $m_*=\min_y m(y)>0$. Since $T$ is a positive Markov operator, it is enough to take $h\ge0$ with $\Norm h_2=1$. Write $h=c+w$, where $c=\E_mh\in[0,1]$, $\E_mw=0$, and $\Norm w_2^2=1-c^2$. Then $\Norm{Tw}_2\le\rho\Norm w_2$ and $\Norm{Tw}_\infty\le m_*^{-1/2}\Norm w_2$.

Choose $\delta\in(0,1/4)$ sufficiently small that $L=\sqrt{2\delta/m_*}/(1-\delta)<1/2$ and $\rho^2(1+L)<1$. Suppose first that $c>1-\delta$. Then $|Tw|\le Lc$. For $2\le q\le3$, Taylor's formula and $|s|\le Lc$ give
\begin{equation}\label{eq:taylor}
(c+s)^q\le c^q+qc^{q-1}s+\frac{q(q-1)}2c^{q-2}(1+L)s^2.
\end{equation}
Indeed, the second derivative along the segment from $c$ to $c+s$ is at most $q(q-1)c^{q-2}(1+L)^{q-2}\le q(q-1)c^{q-2}(1+L)$. Taking expectations, using $\E_\nu Tw=0$, and choosing $q>2$ close enough to $2$ that $(q-1)\rho^2(1+L)\le1$, we obtain
\[
\E_\nu(Th)^q\le c^q+\frac q2c^{q-2}(1-c^2)\le(c^2+1-c^2)^{q/2}=1.
\]
The second inequality follows from convexity of $x\mapsto x^{q/2}$.

Now suppose that $c\le1-\delta$. Since $\E_\nu Tw=0$, we have $\Norm{Th}_2^2\le c^2+\rho^2(1-c^2)\le1-\kappa$, where $\kappa=(1-\rho^2)(2\delta-\delta^2)>0$. Also $\Norm{Th}_\infty\le\Norm h_\infty\le m_*^{-1/2}$. Consequently,
\[
\Norm{Th}_q^q\le\Norm{Th}_\infty^{q-2}\Norm{Th}_2^2\le m_*^{-(q-2)/2}(1-\kappa)\le1
\]
when $q-2>0$ is sufficiently small. Taking the minimum of the permissible exponents for $T$ and $T^*$ establishes \eqref{eq:hyper}.

For completeness, tensorization follows by induction. Consider a function of two independent coordinates. First apply the $L^2\to L^q$ contraction in the first coordinate while keeping the second coordinate fixed. The resulting mixed norm is $L^q$ in the first output and $L^2$ in the second input. By Minkowski's inequality for $q\ge2$, this is bounded by the norm obtained in the reverse order. The contraction in the second coordinate then gives the $L^2$ norm of the original function. Hence $\Norm{(T_1\otimes T_2)h}_q\le\Norm h_2$. Induction proves the assertion for any tensor power, and the same argument applies to the adjoints.
\end{proof}

\begin{lemma}\label{lem:levelone}
Let $Y_1,\ldots,Y_n$ be independent with arbitrary finite distributions, and let $h(Y)\in[0,1]$ have mean $a$. Put $\ell_i(y)=\E[h\mid Y_i=y]-a$ and $W_1=\sum_i\E\ell_i(Y_i)^2$. If $|\ell_i(y)|\le\lambda$ on the support of each coordinate distribution, then
\begin{equation}\label{eq:levelone}
W_1\le I(a)^2+C\lambda.
\end{equation}
The constant is absolute and does not depend on the coordinate distributions.
\end{lemma}
\begin{proof}
Let $L=\sum_i\ell_i(Y_i)$ and $w^2=\E L^2=W_1$. Orthogonality of the first-level components gives $\E[hL]=w^2$. For each real $z$, the pointwise inequality $h(L-z)\le (L-z)_+$ and the identity $\E h=a$ yield $w^2\le az+\E(L-z)_+$. The Wasserstein normal approximation for independent centered summands \cite{CGS11} gives $d_{W_1}(L,wZ)\le Cw^{-2}\sum_i\E|\ell_i(Y_i)|^3\le C\lambda$ when $w>0$. Minimizing $az+\E(wZ-z)_+$ over $z$ now gives $w^2\le wI(a)+C\lambda$. If $w>I(a)$, then $(w-I(a))(w+I(a))\le C\lambda(w+I(a))/w\le2C\lambda$. Otherwise $w^2\le I(a)^2$. The case $w=0$ is immediate.
\end{proof}

\subsection{The posterior diffusion and its energy bounds}
Write $K(y\mid x)=P(y\mid x)$ and $\Sigma_p=\operatorname{diag}(p)-pp^\top$. For each coordinate, let $\mu_t^i$ be an independent finite-type Wright--Fisher diffusion on the simplex over $\cX$, initially equal to $\nu$, with generator $\frac12\sum_{x,x'}(\mu_x\ind_{x=x'}-\mu_x\mu_{x'})\partial_x\partial_{x'}$. The finite-dimensional diffusion with this generator exists on the closed simplex and has continuous paths; see, for example, the Wright--Fisher construction in \cite{EK86}. Each coordinate function $\mu_x$ is harmonic for the generator, so $\mu_t^i$ is a bounded probability-vector martingale. Applying the generator to $\mu_x\mu_{x'}$ yields $\frac{d}{dt}\E\Sigma_{\mu_t^i}=-\E\Sigma_{\mu_t^i}$ and hence
\begin{equation}\label{eq:WFmoments}
\E\mu_t^i=\nu,\qquad \E[\mu_t^i(\mu_t^i)^\top]=\nu\nu^\top+(1-e^{-t})\Sigma_\nu,\qquad \E\Sigma_{\mu_t^i}=e^{-t}\Sigma_\nu.
\end{equation}
The bounded martingale convergence theorem gives an almost-sure limit for every $\mu_t^i$. Since $\E\operatorname{tr}\Sigma_{\mu_t^i}\to0$, the limit is a vertex $e_{X_i}$. Its expectation is $\nu$, so $X_i\sim\nu$; independence of the diffusions makes the $X_i$ independent. Define $\pi_t^i=K^\top\mu_t^i$. For the multi-affine extensions of $f$ and $g$, write $F_t=\widetilde f(\mu_t^1,\ldots,\mu_t^n)$ and $G_t=\widetilde g(\pi_t^1,\ldots,\pi_t^n)$. A multi-affine polynomial in independent continuous martingales is a martingale: It\^o's formula has no diagonal second derivatives and the cross variations of different coordinate diffusions vanish. Both processes are bounded, continuous martingales. Since $0\le F_t,G_t\le1$, their quadratic variations satisfy $\E\langle F\rangle_\infty=\Var(F_\infty)\le1/4$ and $\E\langle G\rangle_\infty=\Var(G_\infty)\le1/4$. Thus the stochastic integrals below are genuine square-integrable martingales after stopping: the derivatives of $J_r$ in either mean lie in $[0,1]$, and its derivative in $r$ is bounded for $0\le r\le\rho<1$. In particular, there is no residual localization error when taking expectations. At the terminal time,
\begin{equation}\label{eq:terminal}
F_\infty=f(X^n),\qquad G_\infty=(T^{\otimes n}g)(X^n),\qquad \E[F_\infty G_\infty]=\E_{P^{\otimes n}}[fg].
\end{equation}

For each $i$ and $y\in\cY$, use the derivative representative
\begin{equation}\label{eq:bi}
b_i(y,t)=\E_{\prod_{j\ne i}\pi_t^j}[g(y,Y_{-i})]-\sum_zm(z)\E_{\prod_{j\ne i}\pi_t^j}[g(z,Y_{-i})].
\end{equation}
Thus $\sum_y m(y)b_i(y,t)=0$ and $b_i$ does not depend on $\mu_t^i$. For $i\ne j$, let $D_{ij}(y,z,t)$ be the mixed derivative that is centered separately in $y$ and $z$ under $m$. It depends only on $\mu_t^\ell$ with $\ell\notin\{i,j\}$, and $b_i(\pi_j)-b_i(m)=D_{ij}(\pi_j-m)$ with the other posterior vectors held fixed.

\begin{lemma}\label{lem:energies}
There is $C_P<\infty$ such that, uniformly over $n$ and $t\ge0$,
\begin{equation}\label{eq:energies}
\sum_i\E\Norm{b_i(t)}_2^2\le C_P,\qquad \sum_{i\ne j}\E\Norm{D_{ij}(t)}_F^2\le C_P.
\end{equation}
\end{lemma}
\begin{proof}
Maximal correlation gives $K^\top\Sigma_\nu K\preceq\rho^2\Sigma_m$. More explicitly, setting $u_t=1-e^{-t}$, the posterior covariance equals $\E[(\pi_t^i-m)(\pi_t^i-m)^\top]=u_tK^\top\Sigma_\nu K\preceq u_t\rho^2\Sigma_m$. Thus averaging an Efron--Stein component against the posterior in any centered coordinate contracts its squared norm by at most $u_t\rho^2\le\rho^2$. To see the orthogonality, write $g=\sum_S g_S$ in the product $m$-orthogonal decomposition. After averaging in the coordinates outside $i$, each contribution to $b_i$ contains one centered factor $\pi_t^j-m$ for every $j\in S\setminus\{i\}$. Contributions indexed by distinct $S$ are orthogonal after averaging over the independent posterior vectors. Two derivatives similarly leave the factors in $S\setminus\{i,j\}$. Their second moments contribute at most $\rho^{2(|S|-1)}$ and $\rho^{2(|S|-2)}$, respectively, up to fixed alphabet norm-equivalence constants. Hence
\begin{equation}\label{eq:energiesproof}
\sum_i\E\Norm{b_i}_2^2\le C_P\sum_S|S|\rho^{2(|S|-1)}\Norm{g_S}_2^2,\quad \sum_{i\ne j}\E\Norm{D_{ij}}_F^2\le C_P\sum_S|S|(|S|-1)\rho^{2(|S|-2)}\Norm{g_S}_2^2.
\end{equation}
The sums are bounded since $\sum_{S\ne\varnothing}\Norm{g_S}_2^2=\Var(g)\le1/4$. To justify the orthogonality without assuming that the posterior vectors have independent letter entries, fix an orthonormal basis $1,\psi_1,\ldots,\psi_{|\cY|-1}$ of $L^2(m)$. The random variables $\pi_t^j\cdot\psi_a$ have mean zero. Their covariance matrix is bounded by $u_t\rho^2$ times the identity by the preceding posterior covariance estimate. Products of such centered factors indexed by different coordinate sets are orthogonal under the independent diffusions. Differentiating with respect to $y_i$ removes one factor, and differentiating with respect to both $y_i,y_j$ removes two. This gives exactly the powers and combinatorial factors in \eqref{eq:energiesproof}. The equivalence of the Euclidean and weighted letter norms depends only on $m$. The separate $m$-centering of $D_{ij}$ is essential: it makes the mixed derivative depend only on coordinates outside $\{i,j\}$.
\end{proof}

Set $u_t=1-e^{-t}$ and
\begin{equation}\label{eq:beta}
\beta_t=\frac{\rho^2e^{-t}}{1-\rho^2u_t},\qquad \xi_i=K^\top\Sigma_{\mu_t^i}K-\beta_t\Sigma_{\pi_t^i},\qquad \xi_i'=\xi_i-\E\xi_i.
\end{equation}
By \eqref{eq:WFmoments}, writing $S=K^\top\Sigma_\nu K$, we have $\E\xi_i=\frac{e^{-t}}{1-\rho^2u_t}(S-\rho^2\Sigma_m)\preceq0$. In particular the mean of this matrix error is nonpositive in every direction.

\begin{lemma}\label{lem:selfaverage}
Let $\Pi_\lambda$ denote Euclidean projection onto the ball of radius $\lambda\le1$ in $\R^{\cY}$, and put
\begin{equation}\label{eq:Z}
Z_t=\sum_i\bigl(\Pi_\lambda b_i(t)\bigr)^\top\xi_i'(t)\bigl(\Pi_\lambda b_i(t)\bigr).
\end{equation}
Then $\E Z_t=0$ and
\begin{equation}\label{eq:selfaverage}
\E Z_t^2\le C_P\lambda^2e^{-t},\qquad \E|Z_t|\le C_P\lambda e^{-t/2}.
\end{equation}
\end{lemma}
\begin{proof}
Since $\Pi_\lambda b_i$ is independent of $\mu_t^i$ and $\E\xi_i'=0$, each summand has mean zero. The covariance calculation below uses $\E\Norm{\xi_i'}_F^2\le C_Pe^{-t}$, which follows from $\E\operatorname{tr}\Sigma_{\mu_t^i}=O_P(e^{-t})$, $\Norm{\Sigma_{\mu}}_F^2\le\operatorname{tr}\Sigma_\mu$, and $\beta_t=O_P(e^{-t})$. The diagonal contribution to $\E Z_t^2$ is at most $C_Pe^{-t}\sum_i\E\Norm{\Pi_\lambda b_i}^4\le C_P\lambda^2e^{-t}$ by Lemma~\ref{lem:energies}.

For $i\ne j$, condition on all diffusion coordinates except $\mu_t^i,\mu_t^j$. Write $Q_i=(\Pi_\lambda b_i)^\top\xi_i'(\Pi_\lambda b_i)$, and let $Q_i^0$ be obtained by replacing $\pi_t^j$ by $m$ in $b_i$. Define $Q_j^0$ symmetrically. Expand the mixed covariance as
\[
\begin{aligned}
\E[(Q_i-Q_i^0)(Q_j-Q_j^0)\mid\mu_{-ij}]={}&\E[Q_iQ_j\mid\mu_{-ij}]-\E[Q_i^0Q_j\mid\mu_{-ij}]\\
&-\E[Q_iQ_j^0\mid\mu_{-ij}]+\E[Q_i^0Q_j^0\mid\mu_{-ij}].
\end{aligned}
\]
The last three expectations vanish separately. Indeed, $Q_i^0$ depends only on $\mu_t^i$ and $\mu_{-ij}$; after fixing those variables, $Q_j$ has mean zero as a function of $\mu_t^j$ because its derivative does not depend on $\mu_t^j$ and $\E\xi_j'=0$. Hence $\E[Q_i^0Q_j\mid\mu_{-ij}]=0$. The symmetric argument shows $\E[Q_iQ_j^0\mid\mu_{-ij}]=0$. Finally, $Q_i^0$ and $Q_j^0$ are independent after conditioning on $\mu_{-ij}$, and each is centered, so $\E[Q_i^0Q_j^0\mid\mu_{-ij}]=0$. Consequently $\E[Q_iQ_j\mid\mu_{-ij}]=\E[(Q_i-Q_i^0)(Q_j-Q_j^0)\mid\mu_{-ij}]$. The projection is 1-Lipschitz, and $v\mapsto vv^\top$ is $2\lambda$-Lipschitz in Frobenius norm on this ball. Therefore
\begin{equation}\label{eq:diffQi}
|Q_i-Q_i^0|\le2\lambda\Norm{\xi_i'}_F\Norm{D_{ij}}_F\Norm{\pi_t^j-m}_2,
\end{equation}
with the symmetric bound for $Q_j-Q_j^0$. The tensor $D_{ij}$ is fixed under the conditional expectation. After multiplying the two bounds, the factor depending on $\mu_t^i$ is $\Norm{\xi_i'}_F\Norm{\pi_t^i-m}_2$, and the corresponding factor for $j$ is independent of it. Each has expectation at most $C_Pe^{-t/2}$ by Cauchy--Schwarz, because $\E\Norm{\xi_i'}_F^2=O_P(e^{-t})$ and $\E\Norm{\pi_t^i-m}_2^2=O_P(1)$. Therefore
\begin{equation}\label{eq:cross}
\bigl|\E[Q_iQ_j\mid\mu_{-ij}]\bigr|\le C_P\lambda^2e^{-t}\Norm{D_{ij}}_F^2.
\end{equation}
Summing over $i\ne j$ and using Lemma~\ref{lem:energies} completes the variance bound. The first-moment bound follows by Cauchy--Schwarz.
\end{proof}

\subsection{Derivative propagation and the Bellman comparison}
\begin{lemma}\label{lem:propagate}
Under the low-degree influence hypothesis in Theorem~\ref{thm:comparison},
\begin{equation}\label{eq:propagate}
\Pbb\left[\sup_{t\ge0}\max_i\Norm{b_i(t)}_2>C_P\lambda\right]\le C_P\left(\frac{d\tau^{q/2-1}}{\lambda^q}+\frac{(d+1)\rho^{2d}}{\lambda^2}\right).
\end{equation}
\end{lemma}
\begin{proof}
For each fixed $(i,y)$, $b_i(y,t)$ is a bounded martingale. Its terminal value is $T^{\otimes(n-1)}h_{i,y}(X_{-i})$, where $h_{i,y}(Y_{-i})=g(y,Y_{-i})-\E_{z\sim m}g(z,Y_{-i})$. Split this function into Efron--Stein degrees below $d$ and at least $d$. Because one coordinate of $g$ has been fixed, these correspond to degrees at most $d$ and greater than $d$ in the original function $g$, respectively. The low-degree part satisfies $\Norm{h_{i,y}^{<d}}_2^2\le\Inf_i^{\le d}(g)/m_{\min}$. The tensorized $2\to q$ estimate and Doob's inequality therefore give a sum of failure probabilities bounded by $C_P\lambda^{-q}\sum_i(\Inf_i^{\le d}(g))^{q/2}\le C_Pd\tau^{q/2-1}/\lambda^q$. For clarity, $\sum_i\Inf_i^{\le d}(g)=\sum_{1\le |S|\le d}|S|\|g_S\|_2^2\le d\Var(g)\le d/4$. Hence $\sum_i(\Inf_i^{\le d}(g))^{q/2}\le\tau^{q/2-1}d/4$, which proves the stated low-degree term after applying Doob's $L^q$ inequality. For the high-degree part, a component of $g$ supported on $S$ with $|S|>d$ leaves $|S|-1$ centered coordinates after one letter is fixed. Their conditional expectation is contracted by at most $\rho^{|S|-1}$. Doob's $L^2$ inequality gives a bound proportional to $\lambda^{-2}\sum_{|S|>d}|S|\rho^{2(|S|-1)}\|g_S\|_2^2$. This is at most $C_P(d+1)\rho^{2d}/\lambda^2$, after enlarging $C_P$ to handle the initial values of $d$ and using exponential decay in $|S|$. A union bound over the fixed letter alphabet finishes the proof. The estimate controls all letters, including those absent from the current posterior support.
\end{proof}

\begin{proof}[Proof of Theorem~\ref{thm:comparison}]
Fix $\theta,\lambda,T_0$. If either mean lies outside $[\theta,1-\theta]$, the elementary Fr\'echet bounds give $\E[fg]\le J_\rho(\E f,\E g)+\theta$. Otherwise define the normalized instantaneous rates
\begin{equation}\label{eq:ABC}
A_t=\frac{d\langle F\rangle_t/dt}{I(F_t)^2},\qquad B_t=\frac{d\langle G\rangle_t/dt}{I(G_t)^2},\qquad C_t=\frac{d\langle F,G\rangle_t/dt}{I(F_t)I(G_t)}.
\end{equation}
The quadratic-variation Cauchy--Schwarz inequality gives $C_t\le\sqrt{A_tB_t}$. For $0\le r\le\rho$, $J_r(a,b)\ge ab$ because $\partial_rJ_r=\varphi_r\ge0$ and $J_0=ab$. The bound $J_r(a,b)\ge a+b-1$ follows from the Fr\'echet inequality for two events. Moreover, $0\le\partial_aJ_r,\partial_bJ_r\le1$. Its first and second derivatives imply
\begin{equation}\label{eq:bellman}
\bigl[dJ_{r_t}(F_t,G_t)\bigr]_{dt}=\varphi_{r_t}(z_t,w_t)\left(\dot r_t+C_t-\frac{r_t}{2}(A_t+B_t)\right)dt,
\end{equation}
where $z_t=\Phi^{-1}(F_t)$, $w_t=\Phi^{-1}(G_t)$, and $\varphi_r$ is the bivariate normal density. Completing the square in $\sqrt{A_t}$ gives $C_t-r_t(A_t+B_t)/2\le (1-r_t^2)B_t/(2r_t)$ for $r_t>0$.

Choose $r_t$ by
\begin{equation}\label{eq:clock}
r_t^2=\frac{\rho^2e^{-t}}{1-\rho^2+\rho^2e^{-t}}=\beta_t,\qquad \dot r_t=-\frac{\beta_t(1-r_t^2)}{2r_t},\qquad r_0=\rho.
\end{equation}
Before applying the level-one lemma, note the exact quadratic-variation identity
\[
\begin{aligned}
\frac{d\langle G\rangle_t}{dt}&=\sum_i b_i(t)^\top K^\top\Sigma_{\mu_t^i}K b_i(t)\\
&=\beta_t\sum_i\Var_{\pi_t^i}(b_i(Y_i,t))+\sum_i b_i(t)^\top\xi_i(t)b_i(t).
\end{aligned}
\]
For the function $g$ under the product distribution $\bigotimes_i\pi_t^i$, its first-level component in coordinate $i$ is exactly $\ell_i(y)=b_i(y,t)-\sum_z\pi_t^i(z)b_i(z,t)$. Apply Lemma~\ref{lem:propagate} with argument $\lambda/C_P$ and enlarge its source-dependent constant; this controls the event $\sup_t\max_i\Norm{b_i(t)}_2>\lambda$ by the same right-hand side as in that lemma. Stop when $F_t$ or $G_t$ first hits $\theta$ or $1-\theta$, when $\max_i\Norm{b_i(t)}_2$ first reaches $\lambda$, or at $T_0$. Before this stop, $|\ell_i(y)|\le2\lambda$. Apply Lemma~\ref{lem:levelone} to this product distribution. The negative semidefinite mean of $\xi_i$ can be discarded, and Lemma~\ref{lem:selfaverage} uses the projection radius $\lambda$, which agrees with $b_i$ before the derivative stop. We obtain
\begin{equation}\label{eq:Bbound}
B_t\le\beta_t+\frac{C_P\beta_t\lambda+Z_t}{I(G_t)^2},
\end{equation}
where $Z_t$ is the centered truncated sum from Lemma~\ref{lem:selfaverage}.

The density factor in \eqref{eq:bellman} gives an improvement over a uniform density bound. For $r\in[0,\rho]$,
\begin{equation}\label{eq:density}
\varphi_r(z,w)=\frac{I(\Phi(w))}{\sqrt{1-r^2}}\varphi\left(\frac{z-rw}{\sqrt{1-r^2}}\right)\le C_P I(\Phi(w)).
\end{equation}
Before the value stop, $I(G_t)\ge c\theta$, so $\varphi_{r_t}(z_t,w_t)/I(G_t)^2\le C_P/\theta$. Combining \eqref{eq:bellman}--\eqref{eq:density}, the expected positive drift up to the stopping time is at most
\begin{equation}\label{eq:integral}
\frac{C_P\lambda}{\theta}\int_0^{T_0}\frac{\beta_t(1-r_t^2)}{2r_t}\,dt+\frac{C_P}{\theta}\int_0^{T_0}\frac{\E|Z_t|(1-r_t^2)}{2r_t}\,dt\le\frac{C_P\lambda(1+T_0)}{\theta}.
\end{equation}
The first integral equals $\rho-r_{T_0}$. For the second, Lemma~\ref{lem:selfaverage} gives $\E|Z_t|\le C_P\lambda e^{-t/2}$ and \eqref{eq:clock} gives $r_t\ge\rho e^{-t/2}$. Optional stopping is applied only before the value boundary, where the derivatives of $J_r$ are bounded.

At a value-boundary stop, the conditional terminal product is at most $\min(F_t,G_t)$. If $F_t$ or $G_t$ equals $\theta$, the minimum is at most $\theta$. If one equals $1-\theta$, then $\min(F_t,G_t)\le F_t+G_t-1+\theta$. Because $J_r(a,b)\ge\max(0,a+b-1)$, both cases give an upper bound $J_{r_t}(F_t,G_t)+\theta$. At a derivative stop, a loss of at most one is charged to the probability in Lemma~\ref{lem:propagate}. At the deterministic endpoint, $J_{r_{T_0}}(F_{T_0},G_{T_0})\ge F_{T_0}G_{T_0}$. The remaining terminal covariance is small even though no influence assumption is made on $f$. Indeed,
\begin{equation}\label{eq:terminalerr}
\E(G_\infty-G_{T_0})^2=\sum_{S\ne\varnothing}(1-u_{T_0}^{|S|})\Norm{(T^{\otimes n}g)_S}_2^2\le e^{-T_0}\sum_{S\ne\varnothing}|S|\rho^{2|S|}\Norm{g_S}_2^2\le C_Pe^{-T_0}.
\end{equation}
On the event that the stop equals $T_0$, optional sampling implies $\E[(F_\infty-F_{T_0})(G_\infty-G_{T_0})\mid\mathcal F_{T_0}]$ is the difference between the conditional terminal product and $F_{T_0}G_{T_0}$. Its expectation, even after multiplying by the $\mathcal F_{T_0}$-measurable endpoint indicator, has absolute value at most $\|F_\infty-F_{T_0}\|_2\|G_\infty-G_{T_0}\|_2=O_P(e^{-T_0/2})$. Here the first factor is at most one and the second is controlled by the preceding display. Combining the stopped drift estimate, the boundary cases, and the derivative-stop probability proves \eqref{eq:main-comparison}.
\end{proof}

\section{Binary regularity and polynomial approximation}\label{sec:binary-regularity}
The common-coordinate regularity step uses the mean conditional influence of Alice's function. It is independent of Bob's observed prefix.

\begin{lemma}\label{lem:regularity}
For every $d\ge2$ and $s\in(0,1)$, there is a fixed set $H\subseteq[n]$ with $|H|\le ed/s$ such that, for every $i\notin H$,
\begin{equation}\label{eq:regularity}
\E_{X_H}\Inf_i^{\le d}(f_{X_H})\le s.
\end{equation}
\end{lemma}
\begin{proof}
Let $R_\gamma$ denote the resampling noise operator, which multiplies Efron--Stein degree $k$ by $\gamma^k$. Define $V(H)=\E_{X_H}\Norm{R_\gamma f_{X_H}}_2^2\le1$. For every $i\notin H$, the Efron--Stein decomposition and the identity for conditioning on one coordinate give
\[
V(H\cup\{i\})-V(H)=(1-\gamma^2)\E_{X_H}\sum_{S\ni i}\gamma^{2(|S|-1)}\Norm{(f_{X_H})_S}_2^2.
\]
Take $\gamma^2=1-1/d$. Since $(1-1/d)^{d-1}\ge e^{-1}$, revealing any coordinate with $\E_{X_H}\Inf_i^{\le d}(f_{X_H})>s$ increases $V$ by at least $s/(ed)$. Reveal such coordinates until none remain. Since $V\le1$, at most $ed/s$ coordinates are revealed.
\end{proof}

The next lemma bounds deletion after averaging over the correlated prefixes. This allows Bob to choose the deleted coordinates using his own prefix alone.

\begin{lemma}\label{lem:deletion}
Let $H$ satisfy \eqref{eq:regularity}, and let $\tau\in(0,1)$. For each $y\in\cY^H$, put $B_y=\{i\notin H:\Inf_i^{\le d}(g_y)>\tau\}$, and let $\bar g_y$ be obtained by averaging $g_y$ over the coordinates in $B_y$. Then $\E\bar g_y=\E g_y$ and $\max_{i\notin H}\Inf_i^{\le d}(\bar g_y)\le\tau$. Moreover, with $c=\frac12-\frac1q>0$,
\begin{equation}\label{eq:deletion}
\left|\E[f(X^n)g(Y^n)]-\E_{X_H,Y_H}\E[f_{X_H}(X_{H^c})\bar g_{Y_H}(Y_{H^c})]\right|\le C_P\left(\frac{d\,s^c}{\sqrt\tau}\right)^{1/2}+\frac{\rho^{d+1}}4.
\end{equation}
\end{lemma}
\begin{proof}
For fixed prefixes $x,y$, write $a_i(x)=\Inf_i^{\le d}(f_x)$, $b_i(y)=\Inf_i^{\le d}(g_y)$, and $E_i=\{y:b_i(y)>\tau\}$. Since averaging over $B_y$ removes precisely the Efron--Stein components meeting $B_y$, orthogonality, Cauchy--Schwarz, and maximal correlation show that the absolute change in the conditional product is at most
\[
\frac12\left(\sum_{i\notin H}a_i(x)\ind_{E_i}(y)\right)^{1/2}+\frac{\rho^{d+1}}4.
\]
For the low-degree term, the first factor in the Cauchy--Schwarz estimate is at most $\big(\sum_{i\in B_y}a_i(x)\big)^{1/2}$ and the second is at most $\sqrt{\Var(g_y)}\le1/2$. For the high-degree term, tensorized maximal correlation gives the displayed bound. Jensen's inequality therefore reduces the average loss to
\begin{equation}\label{eq:jensen}
\frac12\left(\sum_{i\notin H}\E[a_i(X_H)\ind_{E_i}(Y_H)]\right)^{1/2}+\frac{\rho^{d+1}}4.
\end{equation}
Put $p_i=\E_{X_H}a_i(X_H)\le s$ and $r_i=\E_{Y_H}b_i(Y_H)$. Since $0\le a_i\le1/4$, H\"older's inequality and the tensorized forward estimate in Lemma~\ref{lem:hyper} imply
\[
\begin{aligned}
\E[a_i(X_H)\ind_{E_i}(Y_H)]&=\E_{\nu^{\otimes H}}[a_i(X_H)(T^{\otimes H}\ind_{E_i})(X_H)]\\
&\le\Norm{a_i}_{L^{q'}(\nu^{\otimes H})}\Norm{T^{\otimes H}\ind_{E_i}}_{L^q(\nu^{\otimes H})}\\
&\le C_Pp_i^{1-1/q}\Pbb(Y_H\in E_i)^{1/2}.
\end{aligned}
\]
Here $T^{\otimes H}$ maps Bob's indicator function on $\cY^H$ to a function of Alice's $X_H$, while $q'=q/(q-1)$. Markov's inequality gives $\Pbb(Y_H\in E_i)\le r_i/\tau$. Since $1-1/q=1/2+c$, we have
\[
\sum_{i\notin H}\E[a_i(X_H)\ind_{E_i}(Y_H)]\le\frac{C_Ps^c}{\sqrt\tau}\sum_{i\notin H}\sqrt{p_ir_i}\le\frac{C_Pd\,s^c}{\sqrt\tau}.
\]
The last inequality follows from Cauchy--Schwarz and $\sum_i p_i,\sum_i r_i\le d/4$, which hold by Parseval for every conditional function. Substitution into \eqref{eq:jensen} proves \eqref{eq:deletion}. The deleted set depends only on $y$, so this operation is noninteractive.
\end{proof}

\begin{proof}[Proof of Theorem~\ref{thm:sample}]
Suppose first that $0<\rho<1$. Write $v=q/2-1$ and $c=1/2-1/q$. For a sufficiently small error parameter $\varepsilon$, choose
\begin{equation}\label{eq:parameters}
\begin{aligned}
\theta&\asymp_P\varepsilon, & T_0&\asymp\log(1/\varepsilon), & \lambda&\asymp_P\frac{\varepsilon^2}{1+T_0},\\
d&\asymp_P\log(1/\varepsilon), & \tau&\asymp_P\left(\frac{\varepsilon\lambda^q}{d}\right)^{1/v}, & s&\asymp_P\left(\frac{\varepsilon^2\sqrt\tau}{d}\right)^{1/c}.
\end{aligned}
\end{equation}
The constants are chosen so that the five terms in \eqref{eq:main-comparison}, the two terms in \eqref{eq:deletion}, and the deterministic-time error are together at most $C_P\varepsilon$. In particular, $d$ is taken large enough that both $\rho^{d+1}$ and $(d+1)\rho^{2d}/\lambda^2$ are $O_P(\varepsilon)$.

For an arbitrary $n$-sample protocol $(f,g)$, choose $H$ by Lemma~\ref{lem:regularity}. Bob replaces $g_y$ by $\bar g_y$, which preserves its conditional mean and has low-degree influences at most $\tau$. Lemma~\ref{lem:deletion} controls the total change after averaging over the prefixes. Apply Theorem~\ref{thm:comparison} on each paired prefix to obtain, with $a_x=\E f_x$ and $b_y=\E g_y$,
\begin{equation}\label{eq:headcomparison}
\E[f(X^n)g(Y^n)]\le\E[J_\rho(a_{X_H},b_{Y_H})]+C_P\varepsilon\le V_{|H|}(a,b)+C_P\varepsilon.
\end{equation}
The regularity lemma gives $|H|\le ed/s$. Substituting \eqref{eq:parameters}, and absorbing powers of $\log(1/\varepsilon)$, yields
\begin{equation}\label{eq:headbound}
|H|\le C_P\varepsilon^{-A(q)}\operatorname{polylog}(1/\varepsilon),\qquad A(q)=\frac{2q(4q-3)}{(q-2)^2}.
\end{equation}

It remains to implement the Gaussian tail. Choose centered, unit-variance functions $\phi:\cX\to\R$ and $\psi:\cY\to\R$ with $\E[\phi(X)\psi(Y)]=\rho$. On $m_0$ fresh samples, form the normalized sums of $\phi(X)$ and $\psi(Y)$. Their covariance matrix has eigenvalues $1\pm\rho>0$. The two-dimensional convex-set Berry--Esseen theorem of Bentkus \cite{Ben03}, applied after whitening, gives an $O_P(m_0^{-1/2})$ error uniformly over lower rectangles. The one-dimensional marginals satisfy the same bound. Alice and Bob choose threshold tests, randomizing privately at a threshold atom to obtain exactly $a_x$ and $b_y$. Every atom of a marginal normalized sum has probability $O_P(m_0^{-1/2})$, since its distribution function is uniformly close to a continuous Gaussian distribution. Consequently the Gaussian probabilities at the adjacent thresholds differ from the prescribed marginals by $O_P(m_0^{-1/2})$. Since $J_\rho$ is 1-Lipschitz in each argument, the joint output probability differs from $J_\rho(a_x,b_y)$ by $O_P(m_0^{-1/2})$, uniformly over all prefixes and their prescribed means.

Taking $m_0=O_P(\varepsilon^{-2})$ and combining the tail construction with \eqref{eq:headcomparison} proves $M_\infty(a,b)-M_{N(\varepsilon)}(a,b)\le C_P\varepsilon$ for $N(\varepsilon)\le C_P\varepsilon^{-A(q)}\operatorname{polylog}(1/\varepsilon)$ uniformly over $(a,b)$. Absorbing the logarithms into $\varepsilon^{-\delta}$ and inverting gives \eqref{eq:rate}.

For fixed marginals, the attainable values of $\E[fg]$ form an interval. Indeed, the two admissible sets of $[0,1]$-valued functions with prescribed means are convex, their product is connected, and $(f,g)\mapsto\E fg$ is continuous. The two endpoints are $M_n(a,b)$ and $a-M_n(a,1-b)$, the latter by complementing Bob. Thus convergence of both endpoints in the displayed uniform estimate gives the total-variation statement without changing the output marginals.

If $\rho=0$, the source outputs are independent, and every simulable binary law is a product law, available with private randomization. If $\rho=1$, equality in maximal correlation gives a nonconstant common function $Z=\phi(X)=\psi(Y)$ almost surely. Writing $p=\max_z\Pbb(Z=z)<1$, each atom of $Z^n$ has mass at most $p^n$. Partitioning these atoms into four groups with masses approximating any prescribed four-cell distribution gives total-variation error at most $3p^n$. This needs $O_P(\log(1/\varepsilon))$ samples and completes the proof for all finite sources.
\end{proof}

\begin{proof}[Proof of Corollary~\ref{cor:head}]
The Gaussian tail construction gives $V_k(a,b)\le M_\infty(a,b)$. Padding with unused coordinates makes $V_k$ nondecreasing. The bound \eqref{eq:headcomparison} holds for every original protocol with a head of size at most \eqref{eq:headbound}. Thus $M_\infty(a,b)\le V_k(a,b)+C_P\varepsilon$ once $k\ge C_P\varepsilon^{-A(q)}\operatorname{polylog}(1/\varepsilon)$. This proves \eqref{eq:head-rate}. For a finite-sample protocol with marginals $(a_n,b_n)$ and joint mass $c_n$, complementing Bob gives $a_n-M_\infty(a_n,1-b_n)\le c_n\le M_\infty(a_n,b_n)$. Passing to a limit of finite protocols and applying Lemma~\ref{lem:Lipschitz} yields the same inequalities for every law in the closure. Conversely, the finite-protocol admissible strategy space at fixed marginals is connected, and the joint mass is continuous in the two strategies. Thus every intermediate value between the limiting lower and upper endpoints is in the closure. A negative-correlation Gaussian tail is obtained by complementing one threshold. Intermediate values follow by private marginal-preserving randomization toward the product law.
\end{proof}

\subsection{A biased binary source with an explicit exponent}
For the biased source
\begin{equation}\label{eq:exampleP}
P=\begin{pmatrix}1/2&1/8\\[1mm]1/8&1/4\end{pmatrix},\qquad \rho=\frac7{15},
\end{equation}
the common marginal is $\operatorname{Bernoulli}(3/8)$. If $Z=(Y-3/8)/(\sqrt{15}/8)$, then $\E Z=0$, $\E Z^2=1$, $\E Z^3=2/\sqrt{15}$, and $\E Z^4=19/15$. Every one-letter function has the form $h=c+sZ$, and symmetry gives $Th=c+\rho sZ_X$. We can certify the stronger contraction $\|Th\|_5\le\|h\|_2$. If $s\ne0$, put $x=\sqrt{15}\,c/s$. The desired inequality is equivalent to
\[
\frac58\left|x-\frac75\right|^5+\frac38\left|x+\frac73\right|^5\le(x^2+15)^{5/2}.
\]
For $x\notin[-7/3,7/5]$, square both sides and consider the polynomial $P_{\rm out}(x)$ in Appendix~\ref{app:q5}; the corresponding polynomial $P_{\rm mid}(x)$ applies on the middle interval. The calculation there proves both expressions positive by exact rational inequalities. The case $s=0$ is equality. Therefore both $T$ and $T^*$ are $L^2\to L^5$ contractions, since the source is symmetric. We may take $q=5$ in Theorem~\ref{thm:sample} and obtain
\begin{equation}\label{eq:examplebound}
N_P(\varepsilon)\le C_\delta\varepsilon^{-(170/9+\delta)}\qquad\text{for every fixed }\delta>0.
\end{equation}
The exponent reflects the regularity estimate rather than a matching lower bound.

\section{Averaged regularity for general finite outputs}\label{sec:vector-regularity}
We now turn to arbitrary finite output alphabets. Both players use the same set of head coordinates, but they average over exceptional remaining coordinates using only their respective heads. Unlike the binary proof, the general theorem does not use Gaussian threshold extremality.

For $f:\mathcal X^n\to\Delta_k$, we use the vector Efron--Stein decomposition. The vector low-degree influence uses the Euclidean norm on the output coordinates:
\[
\Inf_i^{\le d}(f)=\sum_{\substack{S\ni i\\|S|\le d}}\E\|f_S\|_2^2.
\]
Since $\E\|f\|_2^2\le1$, the sum of these influences over all coordinates is at most $d$. We use the same convention on Bob's space.

\begin{lemma}\label{lem:vector-regularity}
Given $d\ge2$ and $0<s<1$, every pair of simplex-valued functions $f:\mathcal X^n\to\Delta_k$ and $g:\mathcal Y^n\to\Delta_m$ has a common fixed coordinate set $H$ with $|H|\le2ed/s$, such that for each $i\notin H$,
\begin{equation}\label{eq:vector-reg}
\E_{X_H}\Inf_i^{\le d}(f_{X_H})\le s,\qquad\E_{Y_H}\Inf_i^{\le d}(g_{Y_H})\le s.
\end{equation}
\end{lemma}
\begin{proof}
Take the resampling noise operator $R_\gamma$ with $\gamma^2=1-1/d$. For each head $H$, define the two potentials $\E_{X_H}\|R_\gamma f_{X_H}\|_2^2$ and $\E_{Y_H}\|R_\gamma g_{Y_H}\|_2^2$, with Euclidean norms on the output vectors. Each potential is at most one. To calculate the increment, decompose each conditional function $f_{X_H}$ into Efron--Stein components indexed by subsets of the remaining coordinates. Averaging after revealing $i$ turns components containing $i$ into a new constant factor in that coordinate. Their noise weights change from $\gamma^{2|S|}$ to $\gamma^{2(|S|-1)}$. Thus the expected increase in the potential is $(1-\gamma^2)\sum_{S\ni i}\gamma^{2(|S|-1)}\|f_S\|_2^2$, averaged over the prefix. For $|S|\le d$, $\gamma^{2(|S|-1)}=(1-1/d)^{|S|-1}\ge e^{-1}$. Therefore an averaged influence exceeding $s$ increases the corresponding potential by at least $s/(ed)$. Reveal such a coordinate on either side until none remains. Since the sum of the two potentials is at most two, the process reveals at most $2ed/s$ coordinates.
\end{proof}

For a prefix $x$, let $A_x=\{i\notin H:\Inf_i^{\le d}(f_x)>\tau\}$, and let $\bar f_x$ be the conditional expectation of $f_x$ over the coordinates in $A_x$. Define $B_y$ and $\bar g_y$ in the same way. Both averaging operations preserve simplex values and conditional output means. In particular, all low-degree influences of $\bar f_x$ and $\bar g_y$ are at most $\tau$.

\begin{lemma}\label{lem:vector-deletion}
If $H$ satisfies \eqref{eq:vector-reg}, $q>2$ satisfies Lemma~\ref{lem:hyper}, and $c=1/2-1/q$, then
\begin{equation}\label{eq:vector-deletion}
\left\|\E[f(X^n)g(Y^n)^\top]-\E[\bar f_{X_H}(X_{H^c})\bar g_{Y_H}(Y_{H^c})^\top]\right\|_1\le C_Pkm\left(\sqrt d\,s^{c/2}\tau^{-1/4}+\rho^{d+1}\right).
\end{equation}
The coordinate deletions are local: Alice's set depends only on $x$, and Bob's only on $y$.
\end{lemma}
\begin{proof}
We first delete Alice's exceptional coordinates, leaving $g$ unchanged. Fix prefixes $x,y$ and output indices $a,b$. Put $h=(f_x)_a$ and $v=(g_y)_b$, and write their Efron--Stein expansions on the unrevealed coordinates. The difference $h-\E[h\mid X_{H^c\setminus A_x}]$ consists precisely of the components indexed by sets meeting $A_x$. The joint expectation of components indexed by different sets is zero. For $|S|\le d$, the conditional expectation operator on the matching $S$ component has norm at most one. Cauchy--Schwarz therefore gives
\[
\left|\sum_{\substack{|S|\le d\\S\cap A_x\ne\varnothing}}\E[(h_S)(v_S)]\right|\le\|h\|_2\left(\sum_{\substack{|S|\le d\\S\cap A_x\ne\varnothing}}\|v_S\|_2^2\right)^{1/2}\le\left(\sum_{i\in A_x}\Inf_i^{\le d}(g_y)\right)^{1/2}.
\]
On degrees greater than $d$, every nonconstant coordinate contracts by at most $\rho$ under the source channel. The tensor-product operator thus has norm at most $\rho^{d+1}$ on that subspace, while $\|h\|_2,\|v\|_2\le1$. The conditional error for this output entry is consequently at most $\bigl(\sum_{i\in A_x}\Inf_i^{\le d}(g_y)\bigr)^{1/2}+\rho^{d+1}$. Jensen's inequality bounds the average of its first term by
\[
\left(\sum_{i\notin H}\E\left[\mathbf1_{\{\Inf_i^{\le d}(f_{X_H})>\tau\}}\Inf_i^{\le d}(g_{Y_H})\right]\right)^{1/2}.
\]
Write $a_i(x)=\Inf_i^{\le d}(f_x)$ and $b_i(y)=\Inf_i^{\le d}(g_y)$, and set $p_i=\E a_i\le s$ and $r_i=\E b_i\le s$. The simplex constraint gives $0\le a_i,b_i\le1$ and $\sum_i p_i,\sum_i r_i\le d$. Let $A_i=\{x:a_i(x)>\tau\}$. The adjoint channel $T^{*\otimes H}$ maps Alice's indicator $\mathbf1_{A_i}$ to a function on Bob's prefix space. Its $L^q$ norm is at most $\Pr(X_H\in A_i)^{1/2}$ by Lemma~\ref{lem:hyper}. Taking the H\"older conjugate exponent $q/(q-1)$ and using $b_i\le1$ gives $\|b_i\|_{q/(q-1)}\le r_i^{1-1/q}$. Together with Markov's inequality this yields
\begin{equation}\label{eq:vector-cross}
\E[\mathbf1_{A_i}(X_H)b_i(Y_H)]\le r_i^{1-1/q}\Pr(X_H\in A_i)^{1/2}\le r_i^{1/2+c}\left(\frac{p_i}{\tau}\right)^{1/2},\qquad c=\frac12-\frac1q.
\end{equation}
Since $r_i\le s$, Cauchy--Schwarz over the remaining coordinates gives
\[
\sum_i\E[\mathbf1_{A_i}(X_H)b_i(Y_H)]\le\frac{s^c}{\sqrt\tau}\sum_i\sqrt{p_ir_i}\le\frac{d\,s^c}{\sqrt\tau}.
\]
It follows that the first deletion changes any scalar output probability by at most $\sqrt d\,s^{c/2}\tau^{-1/4}+\rho^{d+1}$.

We next delete Bob's exceptional coordinates using only his prefix $y$. For fixed $x$, coordinate averaging in $f_x$ deletes Efron--Stein components and therefore never increases any other low-degree influence. Hence the same argument, with the roles exchanged and the forward channel $T^{\otimes H}$ now acting on Bob's indicator, bounds the second deletion by the same quantity. Both maps preserve their respective simplex values and prefix-conditional output means. Finally sum the two errors over all $km$ output entries to obtain \eqref{eq:vector-deletion}.
\end{proof}

\section{Simplex-preserving Gaussian moment lifting}\label{sec:gaussian-lifting}
The next construction is the main link between discrete and Gaussian protocols. It produces a single simplex-valued Gaussian function, rather than separately approximating individual output events. Matching nonmultilinear Hermite coefficients is essential for the resulting Gaussian strategy to preserve the joint output matrix.

Choose orthonormal mean-zero bases on $\mathcal X$ and $\mathcal Y$, of dimensions $r_X=|\mathcal X|-1$ and $r_Y=|\mathcal Y|-1$. Write $C$ for their cross-covariance matrix. Its operator norm is $\rho$. On Gaussian space, use independent standard $r_X$- and $r_Y$-dimensional vectors $Z_i,W_i$, with $\E Z_iW_i^\top=C$ and independence between distinct coordinates. A discrete multilinear Fourier coefficient indexed by at most one nonconstant basis function from each input coordinate is identified with the Gaussian Hermite coefficient having the same first-order basis index. We write $\mathcal E h^{\le d}$ for this embedding, with all other Hermite coefficients set to zero.

\begin{lemma}\label{lem:mixed-hyper}
Fix either input marginal of the source. Consider a product probability space whose independent coordinates have this finite marginal or are standard Gaussian variables. On a finite coordinate, a nonconstant Efron--Stein component has degree one; on a Gaussian coordinate, degree means the usual Hermite degree. There is a constant $C_P$ such that every scalar or Euclidean-vector-valued polynomial $p$ of total mixed degree at most $d$ satisfies
\begin{equation}\label{eq:mixed-hyper}
\bigl(\E\|p\|_2^3\bigr)^{1/3}\le C_P^d\bigl(\E\|p\|_2^2\bigr)^{1/2}.
\end{equation}
The same constant can be used for both input marginals. It is independent of the number of coordinates, the Gaussian dimensions, and the dimension of the output vector.
\end{lemma}
\begin{proof}
Let $\pi$ be one of the two finite marginals, and put $\pi_*=\min_{x:\pi(x)>0}\pi(x)$. For a scalar function $h=a+v$ with $\E_\pi v=0$ and $b^2=\E_\pi v^2$, the resampling operator is $R_t h=a+tv$. Since $\|v\|_\infty\le \pi_*^{-1/2}b$, we have $\E_\pi|v|^3\le\pi_*^{-1/2}b^3$. Taylor's formula for $x\mapsto |x|^3$ gives $|a+z|^3\le|a|^3+3a|a|z+3|a|z^2+|z|^3$. Taking expectations therefore yields
\[
\E_\pi|R_t h|^3\le|a|^3+3|a|t^2b^2+\pi_*^{-1/2}t^3b^3.
\]
For all $a\in\R$ and $b\ge0$, the elementary bound $(a^2+b^2)^{3/2}\ge |a|^3+\frac12|a|b^2+\frac12b^3$ holds. Indeed, if $|a|\le b/2$ the right side is at most $7b^3/8$; otherwise convexity gives $(a^2+b^2)^{3/2}\ge|a|^3+3|a|b^2/2$, which is sufficient. Thus any $0<t\le \min\{6^{-1/2},(\pi_*^{1/2}/2)^{1/3}\}$ satisfies $\|R_t h\|_{L^3(\pi)}\le\|h\|_{L^2(\pi)}$, with no sign restriction on $h$.

Choose $t>0$ small enough for this contraction on both input marginals and for $t\le2^{-1/2}$. The Gaussian Ornstein--Uhlenbeck operator, which acts by $t^j$ on Hermite degree $j$, is an $L^2\to L^3$ contraction for this $t$ by Gaussian hypercontractivity. Tensorizing these one-coordinate contractions gives an $L^2\to L^3$ contraction on every mixed product, with the same $t$; the usual mixed-norm Minkowski argument applies because $3\ge2$. The tensor product operator acts by $t^j$ on every homogeneous mixed-degree-$j$ component. If $p$ has total degree at most $d$, write $p=R_t^{\otimes}h$ by dividing each coefficient of degree $j$ by $t^j$. Orthogonality gives $\|h\|_2\le t^{-d}\|p\|_2$. Hence $\|p\|_3\le t^{-d}\|p\|_2$. For Euclidean-vector-valued $p$, Minkowski's inequality gives $\bigl(\E\|p\|_2^3\bigr)^{2/3}\le\sum_a(\E|p_a|^3)^{2/3}\le t^{-2d}\sum_a\E|p_a|^2$. Taking $C_P=t^{-1}$ proves the assertion.
\end{proof}

\begin{lemma}\label{lem:lifting}
Let $f:\mathcal X^n\to\Delta_k$ satisfy $\max_i\Inf_i^{\le d}(f)\le\tau$, where $d\ge2$ and $0<\tau<1/2$. There exists a measurable $F:(\R^{r_X})^n\to\Delta_k$ such that
\begin{equation}\label{eq:lifting}
\left\|F^{\le d}-\mathcal E f^{\le d}\right\|_{L^2(\mathcal N(0,I_{n r_X});\R^k)}\le C_P^d(d+1)^{C_P}(1+\log k)\tau^{1/8}.
\end{equation}
The truncation on the left includes every Hermite monomial of total degree at most $d$, including powers greater than one within a coordinate block.
\end{lemma}
\begin{proof}
We give the convex-duality argument. Let $\mathcal C$ be the set of degree-at-most-$d$ coefficient vectors of Gaussian $\Delta_k$-valued functions. It is convex. It is compact because the set of bounded simplex-valued functions is weak-* closed and compact in $L^\infty$, while each of the finitely many coefficient maps is weak-* continuous. Its support function in the direction of a vector Gaussian polynomial $p=(p_1,\ldots,p_k)$ is
\[
\sup_{F:\,F(z)\in\Delta_k}\E[F(Z)\cdot p(Z)]=\E\max_{1\le a\le k}p_a(Z).
\]
Indeed, an upper bound follows pointwise from the simplex constraint. Equality follows by assigning all mass to the first index attaining the maximum, which is a measurable choice. By finite-dimensional separation, a coefficient vector lies within Euclidean distance $\delta$ of $\mathcal C$ if and only if its pairing with every unit coefficient vector is bounded by the corresponding support function plus $\delta$. Thus the same inequality must be proved for every polynomial $p$ of $L^2$ norm one.
It therefore suffices to prove, for each degree-$d$ polynomial $p$ with $\|p\|_2=1$,
\begin{equation}\label{eq:support}
\E[f(X^n)\cdot p_{\rm ml}(X^n)]\le\E\max_a p_a(Z)+C_P^d(d+1)^{C_P}(1+\log k)\tau^{1/8},
\end{equation}
where $p_{\rm ml}$ is the discrete multilinear polynomial obtained from the Gaussian coefficients that use only first-order Hermites within each coordinate block.

For each Gaussian input block, let $I_i(p)=\E\|p-\E[p\mid Z_{[n]\setminus\{i\}}]\|_2^2$. Call a block heavy if $I_i(p)>\eta$, and write $H=\{i:I_i(p)>\eta\}$ and $L=[n]\setminus H$. Since $p$ has total Hermite degree at most $d$ and $\|p\|_2=1$, Parseval gives $\sum_iI_i(p)\le d$, so $|H|\le d/\eta$. Define $p_0=\E[p\mid Z_L]$ and $p_h=p-p_0$. All remaining block influences of $p_0$ are at most $\eta$. Conditional Jensen gives $\E\max_a p_a\ge\E\max_a(p_0)_a$.

To control the discarded terms, put $f_0=\E[f\mid X_L]$. Every Hermite monomial of $p_h$ contains a nonconstant feature from at least one heavy block. The same is true of every term of its multilinear projection $(p_h)_{\rm ml}$. Hence $\E[(p_h)_{\rm ml}\mid X_L]=0$ and, because $f_0$ depends only on $X_L$, $\E[f_0\cdot(p_h)_{\rm ml}]=0$. Orthogonality of the vector Efron--Stein components now gives
\[
\bigl\|(f-f_0)^{\le d}\bigr\|_2^2=\sum_{\substack{|S|\le d\\S\cap H\ne\varnothing}}\|f_S\|_2^2\le\sum_{i\in H}\Inf_i^{\le d}(f)\le\frac{d\tau}{\eta}.
\]
Because $(p_h)_{\rm ml}$ has degree at most $d$ and $\|(p_h)_{\rm ml}\|_2\le\|p\|_2=1$, Cauchy--Schwarz yields
\begin{equation}\label{eq:remove-high}
\left|\E f\cdot(p_{\rm ml}-(p_0)_{\rm ml})\right|=\left|\E(f-f_0)\cdot(p_h)_{\rm ml}\right|\le\sqrt{d\tau/\eta}.
\end{equation}

Expand $p_0$ in the orthonormal Hermite basis. Let $H_\alpha(Z_i)$ denote the normalized multivariate Hermite polynomial indexed by $\alpha\in\mathbb N^{r_X}$. At each block $i$, group all $H_\alpha$ with $1\le|\alpha|\le d$ into a feature vector $V_i$. Every Hermite monomial in $p_0$ uses at most one such feature from each block. Thus, with vector coefficients $c_{S,(\alpha_i)_{i\in S}}$, the expansion can be written
\[
p_0=\sum_{S\subseteq[n]}\ \sum_{\substack{|\alpha_i|\ge1\ (i\in S)\\\sum_{i\in S}|\alpha_i|\le d}}c_{S,(\alpha_i)}\prod_{i\in S}H_{\alpha_i}(Z_i).
\]
This is multilinear in the block feature vectors, although not in the original Gaussian coordinates. Its block influence $I_i$ is the sum of squared coefficients of terms meeting block $i$. Consequently $\sum_i I_i\le d\|p_0\|_2^2\le d$. Compare two systems of independent blocks: in the first, every feature is its Gaussian Hermite polynomial; in the second, degree-one features are the orthonormal basis functions on $\mathcal X$, while all higher-degree features are independent centered unit-variance Gaussians. Each block vector has mean zero and covariance the identity. A Gaussian Hermite feature of degree $j$ is a polynomial of ordinary degree $j$. A replacement feature is either a finite-alphabet basis function or an independent standard Gaussian variable. Consequently every polynomial in every hybrid system is a polynomial on a product of finite and Gaussian spaces of total degree at most $d$, when each nonconstant finite coordinate is assigned degree one. Lemma~\ref{lem:mixed-hyper} applies uniformly to all hybrid systems, with $L^3$ cost at most $C_P^d$. Every partial replacement preserves the $L^2$ norms of the coefficient arrays, since the block features are centered and orthonormal in both systems. Neither fact depends on the number of underlying input coordinates.

Use the smooth maximum $\Psi_h(z)=h\log\sum_{a=1}^k e^{z_a/h}$. It satisfies $\max_a z_a\le\Psi_h(z)\le\max_a z_a+h\log k$ and $\|D^3\Psi_h\|\le Ch^{-2}$. In a block-by-block Lindeberg replacement, the first two Taylor terms agree by the matching means and covariances. For the third-order remainder, let $I_i$ be the Efron--Stein influence of block $i$ on $p_0$. To see the moment bound explicitly, write the coordinate-dependent part as $\sum_\alpha V_{i,\alpha}p_{i,\alpha}$, where the block features $V_{i,\alpha}$ are orthonormal in $L^2$. Lemma~\ref{lem:mixed-hyper}, applied to this entire vector-valued degree-$d$ polynomial in the current hybrid system, implies $\E\|\sum_\alpha V_{i,\alpha}p_{i,\alpha}\|_2^3\le C_P^{3d}I_i^{3/2}$. Replacing $C_P$ by a larger source-dependent constant absorbs the third power. Summing over blocks uses
\[
\sum_iI_i^{3/2}\le\sqrt\eta\sum_iI_i\le d\sqrt\eta.
\]
Consequently the two expectations of $\Psi_h(p_0)$ differ by at most $C_P^d(d+1)^{C_P}d\sqrt\eta/h^2$. The factor polynomial in $d$ accounts for the fixed number of source basis directions and Hermite features of order at most $d$ in each block.

In the second system, condition on all degree-one discrete features. Each multilinear product containing a higher-degree replacement Gaussian has conditional mean zero, even if several such features occur in different blocks, because the replacements are independent and centered. Thus the conditional mean of $p_0$ is precisely $(p_0)_{\rm ml}(X^n)$. Applying Jensen to the pointwise maximum gives
\[
\E\max_a(p_0)_a(\text{Gaussian})\ge\E\max_a(p_0)_{{\rm ml},a}(X^n)-C_P^d(d+1)^{C_P}\frac{d\sqrt\eta}{h^2}-h\log k.
\]
The right-hand maximum dominates $\E[f\cdot(p_0)_{\rm ml}]$ because $f$ takes values in $\Delta_k$. Combining with \eqref{eq:remove-high}, and choosing $\eta=\tau^{3/4}$ and $h=\tau^{1/8}$, proves \eqref{eq:support}. The support-function inequality holds uniformly for every coefficient direction of norm one. Applying finite-dimensional separation to the target vector of all degree-at-most-$d$ coefficients therefore produces a coefficient vector in $\mathcal C$ at Euclidean distance at most the right-hand error. By the definition of $\mathcal C$, this vector is realized by a measurable function $F$ taking values in $\Delta_k$. Parseval identifies coefficient-vector distance with the $L^2$ distance in \eqref{eq:lifting}.
\end{proof}

\begin{corollary}\label{cor:transfer}
Suppose both $f:\mathcal X^n\to\Delta_k$ and $g:\mathcal Y^n\to\Delta_m$ have low-degree influences at most $\tau$. There are Gaussian simplex-valued functions $F,G$ on the corresponding $n$ blocks for which
\begin{equation}\label{eq:gaussian-transfer}
\left\|\E[f(X^n)g(Y^n)^\top]-\E[F(Z)G(W)^\top]\right\|_1\le C_Pkm\left(C_P^d(d+1)^{C_P}(1+\log(km))\tau^{1/8}+\rho^{d+1}\right).
\end{equation}
The functions $F$ and $G$ can be selected independently from their respective input functions.
\end{corollary}
\begin{proof}
Apply Lemma~\ref{lem:lifting} to each side using the fixed orthonormal bases. Orthogonality of the product Efron--Stein decomposition gives $\E[f_ag_b]=\sum_S\E[(f_a)_S(g_b)_S]$. The cross-moment matrix on one coordinate is $C$, so on a set $S$ it is the tensor product of $|S|$ copies of $C$. The corresponding degree-one Gaussian Hermite modes have exactly the same cross-moment tensor. Hence the truncated discrete joint correlation equals the corresponding truncated embedded Gaussian correlation, including all output entries. The two lifting errors contribute at most twice the bound in \eqref{eq:lifting} by Cauchy--Schwarz. In either space, the absolute contribution of higher degrees is at most $\rho^{d+1}$, since the degree-$j$ conditional expectation operator has norm at most $\rho^j$. This proves the entrywise estimate. The smoothing error in the support-function argument is $h\log k$ on Alice's side and $h\log m$ on Bob's side, where $h=\tau^{1/8}$. Their sum is $h\log(km)$. The remaining error terms also add, and $2+\log(km)\le2(1+\log(km))$. Thus the factor $1+\log(km)$ in the displayed matrix estimate consistently bounds both lifting errors. Sum over the $km$ entries.
\end{proof}

\section{Gaussian compression and proof of the general theorem}\label{sec:general-proof}
The final step reduces an arbitrary common Gaussian tail to a number of samples polynomial in the desired accuracy. The same projection must be used by every pair of conditional strategies.

The next lemma adapts the normalized random-projection method used in Gaussian dimension reduction \cite{GKR18}. It works with arbitrary measurable simplex-valued functions and with any number of prefix states, because the projection is chosen using the averaged error.

\begin{lemma}\label{lem:gauss-compression}
Let $0<\rho<1$ be fixed, and let $(Z,W)$ consist of any number $n$ of independent standard Gaussian pairs with correlation $\rho$. Fix any probability weights $w_{xy}$ on a finite set of prefix pairs. Given arbitrary measurable $F_x:\R^n\to\Delta_k$ and $G_y:\R^n\to\Delta_m$, and an integer $D\ge C_\rho$, there are functions $\widehat F_x,\widehat G_y$ on $D$ Gaussian coordinates, all obtained using the same deterministic matrix, such that
\begin{equation}\label{eq:gauss-compress}
\left\|\sum_{x,y}w_{xy}\E[F_x(Z)G_y(W)^\top]-\sum_{x,y}w_{xy}\E[\widehat F_x(A)\widehat G_y(B)^\top]\right\|_1\le C_\rho km D^{-1/4},
\end{equation}
where $(A,B)$ are $D$ independent $\rho$-correlated Gaussian pairs. The constant and the required $D$ are independent of $n$ and of the number of prefix states.
\end{lemma}
\begin{proof}
Let $M$ be an $n\times D$ matrix of independent standard Gaussian entries. For $a\ne0$, define $\widehat F_x(a)=F_x(Ma/\|a\|_2)$ and $\widehat G_y(a)=G_y(Ma/\|a\|_2)$. This uses the same matrix for all prefix states and preserves the simplex constraint pointwise. It suffices to show that the expected weighted $\ell^1$ error is $O_\rho(kmD^{-1/4})$, because then some deterministic choice of $M$ achieves that bound.

Fix scalar $0\le F,G\le1$, and put
\[
L_M=\E_{A,B}\left[F\left(\frac{MA}{\|A\|_2}\right)G\left(\frac{MB}{\|B\|_2}\right)\right],\qquad L=\E[F(Z)G(W)],
\]
where $(A_j,B_j)_{j=1}^D$ are independent standard Gaussian pairs with correlation $\rho$, and $(Z_i,W_i)_{i=1}^n$ are independent pairs with the same correlation. Write
\[
S_1=\frac1D\sum_{j=1}^D A_j^2,\qquad S_2=\frac1D\sum_{j=1}^D B_j^2,\qquad S_{12}=\frac1D\sum_{j=1}^D A_jB_j,\qquad R=\frac{S_{12}}{\sqrt{S_1S_2}}.
\]
Conditionally on $A,B$, the $n$ rows of $(MA/\|A\|_2,MB/\|B\|_2)$ are independent Gaussian pairs with correlation $R$. Direct Gaussian moment calculations give
\[
\E(S_1-1)^2=\E(S_2-1)^2=\frac2D,\qquad \E(S_{12}-\rho)^2=\frac{1+\rho^2}{D}.
\]
On the event $|S_1-1|,|S_2-1|\le1/2$, the map $(a,b)\mapsto(ab)^{-1/2}$ is $2$-Lipschitz in each argument on $[1/2,3/2]^2$, and $1/\sqrt{S_1S_2}\le2$. Hence
\[
|R-\rho|\le2|S_{12}-\rho|+2\rho(|S_1-1|+|S_2-1|).
\]
The complement has probability at most $4e^{-cD}$ by the elementary chi-square tail bound, and $|R-\rho|\le2$ everywhere. Therefore
\begin{equation}\label{eq:sample-correlation}
\E|R-\rho|\le\frac{2\sqrt{1+\rho^2}+4\rho\sqrt2}{\sqrt D}+8e^{-cD}\le C D^{-1/2}.
\end{equation}
Here and below, $c>0$ denotes a numerical constant unless given a $\rho$ subscript. The random variables $A_jB_j-\rho$, $A_j^2-1$, and $B_j^2-1$ have uniformly bounded subexponential norms. Bernstein's inequality thus yields, for every $0<t\le1/4$,
\[
\Pr\bigl(\max\{|S_1-1|,|S_2-1|,|S_{12}-\rho|\}>t\bigr)\le6e^{-ct^2D}.
\]
Take $t=(1-\rho)/16$ and $r_0=(1+\rho)/2$. The preceding event bound implies
\begin{equation}\label{eq:compression-exception}
\Pr(|R|>r_0)\le6e^{-c_\rho D},\qquad c_\rho=c(1-\rho)^2>0.
\end{equation}

For $u\in(-1,1)$, let $L(u)$ be the noise correlation of $F,G$ for $n$ independent Gaussian pairs of correlation $u$. The Hermite expansion and Cauchy--Schwarz give $L(u)=\E F\,\E G+\sum_{j\ge1}u^j\langle F_j,G_j\rangle$ with $\sum_j|\langle F_j,G_j\rangle|\le\|F\|_2\|G\|_2\le1$. For $|u|,|v|\le r<1$,
\begin{equation}\label{eq:rho-Lipschitz}
|L(u)-L(v)|\le|u-v|\sup_{j\ge1}jr^{j-1}\le\frac{|u-v|}{1-r}.
\end{equation}
The last inequality follows from $jr^{j-1}\le\sum_{\ell=0}^{j-1}r^\ell\le(1-r)^{-1}$; in particular, no extraneous factor $1/e$ is valid uniformly in $r$. Taking $r=r_0$ and using $0\le L(u)\le1$, equations \eqref{eq:sample-correlation} and \eqref{eq:compression-exception} show
\[
|\E_ML_M-L|=|\E_RL(R)-L(\rho)|\le\frac{\E|R-\rho|}{1-r_0}+6e^{-c_\rho D}\le C_\rho D^{-1/2}.
\]

We next control the variance over the choice of $M$. Use two independent seed pairs $(A,B)$ and $(A',B')$, independent of $M$, and write $u=A/\|A\|$, $v=B/\|B\|$, $u'=A'/\|A'\|$, $v'=B'/\|B'\|$. Given these four directions, one row of $(Mu,Mv,Mu',Mv')$ has a centered Gaussian law whose covariance is partitioned into blocks
\[
\Sigma_{11}=\begin{pmatrix}1&R\\ R&1\end{pmatrix},\qquad \Sigma_{22}=\begin{pmatrix}1&R'\\ R'&1\end{pmatrix},\qquad
\Sigma_{12}=\begin{pmatrix}\langle u,u'\rangle&\langle u,v'\rangle\\\langle v,u'\rangle&\langle v,v'\rangle\end{pmatrix}.
\]
For any two jointly Gaussian vectors with positive definite marginal covariance matrices, their maximal correlation equals
\begin{equation}\label{eq:gaussian-maxcorr}
\left\|\Sigma_{11}^{-1/2}\Sigma_{12}\Sigma_{22}^{-1/2}\right\|_{\mathrm{op}}.
\end{equation}
Indeed, Gaussian conditional expectation diagonalizes in the Hermite basis. Its norm on centered functions is the largest singular value of the whitened cross-covariance, attained by linear functions. Because the $n$ rows are independent, the conditional expectation operator for all rows is the tensor product of the one-row operators. Its norm on centered functions remains equal to the one-row maximal correlation: a tensor factor supported on one row attains the value, while a term supported on several rows is contracted by a product of singular values not exceeding the largest one.

On $|R|,|R'|\le r_0$, the least eigenvalues of $\Sigma_{11},\Sigma_{22}$ are at least $1-r_0=(1-\rho)/2$, so the maximal correlation is at most $\|\Sigma_{12}\|_{\mathrm{op}}/(1-r_0)$. The two seed pairs are independent, and each of the four entries of $\Sigma_{12}$ has second moment $1/D$ by spherical symmetry. Therefore
\begin{equation}\label{eq:compression-cross}
\E\|\Sigma_{12}\|_{\mathrm{op}}\le\left(\E\|\Sigma_{12}\|_F^2\right)^{1/2}=\frac2{\sqrt D}.
\end{equation}
Let $X=F(Mu)G(Mv)$ and $X'=F(Mu')G(Mv')$. Both belong to $[0,1]$. Conditionally on the directions, their covariance under $M$ is bounded by the Gaussian maximal correlation times $\sqrt{\Var_M(X)\Var_M(X')}\le1/4$. The exceptional event has probability at most $12e^{-c_\rho D}$ by \eqref{eq:compression-exception}. Thus
\[
\E_{u,v,u',v'}\bigl|\operatorname{Cov}_M(X,X'\mid u,v,u',v')\bigr|\le\frac{1}{4(1-r_0)}\E\|\Sigma_{12}\|_{\mathrm{op}}+12e^{-c_\rho D}\le C_\rho D^{-1/2}.
\]
It remains to justify that this conditional covariance is exactly the term governing $\Var_M L_M$. Given $(u,v)$, the conditional mean of $X$ under $M$ is $L(R)$; similarly, that of $X'$ is $L(R')$. The two direction pairs are independent. Hence $\E[L(R)L(R')]=(\E L(R))^2=(\E_M L_M)^2$, and averaging the preceding covariance gives $\Var_ML_M\le C_\rho D^{-1/2}$. It follows that
\begin{equation}\label{eq:compression-final}
\E_M|L_M-L|\le\sqrt{\Var_ML_M}+|\E_ML_M-L|\le C_\rho D^{-1/4}.
\end{equation}
Since $1-r_0=(1-\rho)/2$ and $c_\rho$ is proportional to $(1-\rho)^2$, the constants above may be chosen to grow at most by a fixed power of $(1-\rho)^{-1}$. In particular, they are not uniform as $\rho\uparrow1$.

Finally, apply \eqref{eq:compression-final} to each entry of each output matrix $F_xG_y^\top$. Average with the given prefix weights $w_{xy}$, sum over the $km$ entries, and use the triangle inequality. The expected $\ell^1$ error is at most $C_\rho kmD^{-1/4}$, irrespective of the number of prefix pairs. Fix one matrix $M$ realizing this bound. Since the same $M$ was used in every entry and every prefix pair, the resulting response functions form one valid noninteractive Gaussian protocol.
\end{proof}

\begin{lemma}\label{lem:scalarization}
Let $(Z_i,W_i)$ be Gaussian blocks with identity marginal covariances and cross-covariance $C$, where $\|C\|_{\mathrm{op}}=\rho>0$. Any protocol using $n$ such block pairs, including arbitrary simplex-valued functions and any fixed prefix states, is equivalent to a protocol using $n\operatorname{rank}(C)$ scalar $\rho$-correlated Gaussian pairs and private Gaussian randomization. The private randomization can be averaged into simplex-valued deterministic response functions.
\end{lemma}
\begin{proof}
Take a singular-value decomposition $C=U\operatorname{diag}(\sigma_1,\ldots,\sigma_t)V^\top$, with $0\le\sigma_j\le\rho$ and $t=\operatorname{rank}(C)$. For the $j$th singular direction let $(A_j,B_j)$ have correlation $\rho$. Alice uses $A_j$; Bob uses $(\sigma_j/\rho)B_j+(1-\sigma_j^2/\rho^2)^{1/2}Z_j'$, where $Z_j'$ is private standard Gaussian. Unmatched directions are filled with private standard Gaussians, and each side then applies the orthogonal change of basis. This exactly produces the required joint Gaussian block law. Conditional expectations over the independent private Gaussian variables turn randomized responses into simplex-valued functions of the scalar correlated input, without changing the joint output law.
\end{proof}

\subsection{Proof of the general sample theorem}
\begin{proof}[Proof of Theorem~\ref{thm:general}]
We first assume $0<\rho<1$ and fix a protocol $f:\mathcal X^n\to\Delta_k$, $g:\mathcal Y^n\to\Delta_m$. For $d\ge2$ and $0<s,\tau<1$, apply Lemma~\ref{lem:vector-regularity}, then locally delete high-influence coordinates as in Lemma~\ref{lem:vector-deletion}. The first operation uses at most $2ed/s$ head coordinates. The resulting conditional functions have low-degree influences at most $\tau$ and change the joint output matrix by at most \eqref{eq:vector-deletion}.

Apply Corollary~\ref{cor:transfer} separately on each paired prefix. The Gaussian functions may vary with the local prefix, but they all live on the same Gaussian coordinate space and have the same cross-covariance $C$. Averaging the comparisons over the joint prefix law preserves the error bound. Lemma~\ref{lem:scalarization} then replaces these Gaussian blocks with scalar $\rho$-correlated Gaussian pairs. Finally Lemma~\ref{lem:gauss-compression} uses one deterministic projection matrix to reduce the common Gaussian tail to $D$ scalar pairs. Combining these steps gives total-variation error at most
\begin{equation}\label{eq:general-errors}
C_Pkm\left(\sqrt d\,s^{c/2}\tau^{-1/4}+C_P^d(d+1)^{C_P}(1+\log(km))\tau^{1/8}+\rho^{d+1}+D^{-1/4}\right),\qquad c=\frac12-\frac1q>0.
\end{equation}
Here $1+\log(km)$ is the joint upper bound for the two log-sum-exp errors $h\log k$ and $h\log m$ in Corollary~\ref{cor:transfer}. No factor depending on the number of prefix states appears.

The scalar Gaussian tail can be implemented with fresh copies of $P$. Choose maximal-correlation functions $\phi(X),\psi(Y)$ of mean zero and variance one, with covariance $\rho$. For each of the $D$ coordinates, use $N$ independent copies to form the normalized pair of sums, then add independent Gaussian noise of variance $\sigma^2$ on both sides. For any measurable set $B\subseteq\R^2$, convolve $\mathbf1_B$ with the centered Gaussian kernel of covariance $\sigma^2I_2$. For every multi-index $\alpha$ with $|\alpha|=3$, the resulting function has $\|\partial^\alpha(\mathbf1_B*\varphi_{\sigma})\|_\infty\le\|\partial^\alpha\varphi_{\sigma}\|_{L^1}\le C\sigma^{-3}$. Replace the $N$ independent bounded source-feature pairs by Gaussian pairs with the same covariance, one summand at a time. The zeroth-, first-, and second-order Taylor terms cancel in expectation. The sum of the absolute third-order remainders is at most $C_P N\cdot N^{-3/2}\sigma^{-3}=C_P/(\sigma^3\sqrt N)$, uniformly over $B$. Taking the supremum over $B$ gives total-variation distance at most this amount between the noisy normalized sum and the similarly smoothed Gaussian pair. The latter differs from the desired nonsmoothed Gaussian pair by $C_P\sigma^2$, because the covariance matrix has smallest eigenvalue $1-\rho>0$. Taking products over $D$ coordinates gives total-variation error at most
\begin{equation}\label{eq:tail-error}
C_PD\left(\sigma^2+\frac1{\sigma^3\sqrt N}\right).
\end{equation}
The downstream Gaussian response functions may be arbitrary measurable simplex-valued maps, since the estimate controls total variation of the entire input distribution.

To make the parameter dependence explicit, put $L=2km/\varepsilon$. Choose $d=\lceil C_P\log L\rceil$ large enough that $\rho^{d+1}\le L^{-4}$. Since $P$ is fixed, the factor $C_P^d(d+1)^{C_P}(1+\log(km))$ is bounded by $L^{B_P}$ for some constant $B_P$. Choose $\tau=L^{-8(B_P+4)}$ and $s=L^{-C'_P}$, where $C'_P$ is sufficiently large that the first two terms of \eqref{eq:general-errors} are at most $\varepsilon/8$ after multiplication by $C_Pkm$. Choose $D=\lceil C_PL^4\rceil$, $\sigma^2=(C_PLD)^{-1}$, and $N=\lceil C_P(LD)^5\rceil$, with constants large enough to make \eqref{eq:tail-error} at most $\varepsilon/8$. The regularization contribution is controlled by the choice of $s$, while the lifting contribution is $O_P(kmL^{-4})$. The Gaussian projection contributes $O_P(km/L)$ and the final finite-source realization contributes $O_P(1/L)$. Increasing source-dependent constants where necessary makes the total error at most $\varepsilon$. Moreover,
\[
|H|\le C_PL^{C'_P+1},\qquad D\le C_PL^4,\qquad DN\le C_PL^{29}.
\]
Taking $A_P\ge\max\{C'_P+1,29\}$ establishes \eqref{eq:general-main}, uniformly over the original protocol length $n$. This is an exponent supplied by the proof, not a claim of optimality; its source dependence is through the hypercontractive exponent and the constants in Gaussian transfer and compression. A limiting target is handled by first approximating it with some finite protocol, then applying this uniform reduction.

If $\rho=0$, Alice and Bob observe independent sources and private randomness simulates every feasible product law exactly. If $\rho=1$, there exists a nonconstant common random variable $Z=h(X)=\ell(Y)$ almost surely. If its largest atom has mass $p<1$, the largest atom of $Z^n$ has mass at most $p^n$. Partitioning these atoms into $km$ groups approximates any desired $k\times m$ law with total-variation error at most $km\,p^n$. The parties can then output the corresponding pair of labels. Taking $n=O_P(\log(km/\varepsilon))$ completes the proof.
\end{proof}

\paragraph{Acknowledgment.} OpenAI GPT-6 and Anthropic Claude Opus assisted with exploratory mathematical derivations, symbolic calculations, proof review, literature searches, and manuscript editing.

\appendix
\section{Exact verification for the biased binary example}\label{app:q5}
The following elementary polynomial certificate completes the $L^2\to L^5$ calculation used in Section~4.
Set
\[
\begin{aligned}
P_{\rm out}(x)&=(x^2+15)^5-\left[\frac58(x-7/5)^5+\frac38(x+7/3)^5\right]^2,\\
P_{\rm mid}(x)&=(x^2+15)^5-\left[-\frac58(x-7/5)^5+\frac38(x+7/3)^5\right]^2.
\end{aligned}
\]
Then $P_{\rm out}(x)>0$ for all $x\in\mathbb R$, and $P_{\rm mid}(x)>0$ whenever $|x|\le7/3$.

\paragraph{Verification.}
Write $P_{\rm out}(x)=\sum_{j=0}^8a_jx^j$ and $P_{\rm mid}(x)=\sum_{j=0}^{10}b_jx^j$. Direct expansion over the rationals gives $a_{2j}>0$ and $a_{2j+1}^2<a_{2j}a_{2j+2}$ for $j=0,1,2,3$. Hence each quadratic expression $\frac12a_{2j}x^{2j}+a_{2j+1}x^{2j+1}+\frac12a_{2j+2}x^{2j+2}$ is nonnegative. Their sum, together with $\frac12(a_0+a_8x^8)$, equals $P_{\rm out}(x)$ and is strictly positive. For the middle polynomial, all its even coefficients are positive, while direct rational expansion gives
\[
b_0-\sum_{j=0}^{4}|b_{2j+1}|\left(\frac73\right)^{2j+1}=\frac{61416553873430377}{123018750000}>0.
\]
Thus $P_{\rm mid}(x)>0$ for $|x|\le7/3$, as required.
\end{document}